\documentclass[conference]{IEEEtran}
\IEEEoverridecommandlockouts

\usepackage{epsfig}
\usepackage{subcaption}
\usepackage{calc}
\usepackage{amssymb}
\usepackage{amstext}
\usepackage{amsmath}
\usepackage{amsthm}
\usepackage{multicol}
\usepackage{multirow}
\usepackage{pslatex}
\usepackage{makecell}
\usepackage{apalike}
\usepackage{xurl}
\usepackage{tabularx}
\usepackage{xcolor}
\usepackage[bottom]{footmisc}
\usepackage{balance}
\usepackage{tikz}

 \newcommand\copyrighttextbottom{%
  \footnotesize Accepted version of Marcel Mraz, Hind Bangui, Bruno Rossi, and Barbora Buhnova. 2023. Adopting the Actor Model for Antifragile Serverless Architectures. In the 18th International Conference on Software and Data Technologies (ICSOFT), July 10-12, 2023, Rome, Italy. SciTePress.}
\newcommand\copyrightnoticebottom{%
\begin{tikzpicture}[remember picture,overlay]
\node[anchor=south,yshift=24pt] at (current page.south) {\fbox{\parbox{\dimexpr1.0\textwidth-\fboxsep-\fboxrule\relax}{\copyrighttextbottom}}};
\end{tikzpicture}%
}  

\begin{document}

\title{Adopting the Actor Model for Antifragile Serverless Architectures}

\author{\IEEEauthorblockN{Marcel Mraz, Hind Bangui, Bruno Rossi, and Barbora Buhnova}

\IEEEauthorblockA{\textit{Faculty of Informatics, Masaryk University}\\
Brno, Czech Republic \\
\{marcelmraz, hind.bangui, brossi, buhnova\}@mail.muni.cz}}

\maketitle

\copyrightnoticebottom
\begin{abstract}
Antifragility is a novel concept focusing on letting software systems learn and improve over time based on sustained adverse events such as failures. The actor model has been proposed to deal with concurrent computation and has recently been adopted in several serverless platforms. In this paper, we propose a new idea for supporting the adoption of supervision strategies in serverless systems to improve the antifragility properties of such systems. We define a predictive strategy based on the concept of stressors (e.g., injecting failures), in which actors or a hierarchy of actors can be impacted and analyzed for systems' improvement. The proposed solution can improve the system's resiliency
in exchange for higher complexity but goes in the direction of building antifragile systems.
\end{abstract}



\section{\uppercase{Introduction}}
Antifragility is an emerging research area aiming to introduce in a software system and its architecture stressors, variation, randomness, and uncertainties to improve over time~\cite{taleb2012antifragile,hind-icsme2022,hind-iotbds22}. Antifragility was introduced by Nassim Taleb in 2012 in his book \textit{“Antifragile: things that gain from disorder”}~\cite{taleb2012antifragile}, explaining the concept as: \emph{"Some things benefit from shocks; they thrive and grow when exposed to volatility, randomness, disorder, and stressors and love adventure, risk, and uncertainty. Yet, despite the ubiquity of the phenomenon, there is no word for the exact opposite of fragile. Let us call it \textbf{antifragile}."} Compared to the resilience concept, understood as the ability to plan and prepare for, absorb, recover from, and adapt to adverse events~\cite{national2012disaster}, antifragility is not only helping a system resist shocks and return to previous levels of operation after recovery, but it helps to gain from shocks and learn at runtime how to increase the adaptability and evolvability. As a result, antifragility is an improved version of classical resilience that helps a system handle a variety of hazards and strengthen the protection and safety of its components and services under unforeseen changes while interacting with other inter-dependable systems. 

In this paper, we leverage the antifragility concept to progress toward building antifragile serverless systems that can gain from unexpected failures and defects. A simple programming model called Function-as-a-Service (FaaS) was adopted in serverless computing. Each task is represented and executed by an independent and stateless function to provide high computation power and reduce latency, all cost-effectively~\cite{taibi2020serverless}.

In previous works~\cite{hind-icsme2022}, the stressor concept was introduced to help a system to explore its fragilities and set-up mechanisms for learning and improving based on adverse events (failures). Likewise, our idea is to focus on supporting the development of antifragile systems. Thus, we adopt the actor model in this work to support creating antifragile serverless systems.

Many studies have focused on enhancing the resilience of actors to fulfill their planned tasks; particularly, they have focused on using three resilience properties \cite{de2020automated,cao2021intelligent}, which are: a)~robustness: the ability to resist faults/crises, b)~recoverability: the ability to recover from faults/crises rapidly and back to an original condition, and c)~reliability: the ability to fulfill tasks under stress or fault conditions. However, the recoverability policy options might be exhausted due to the hazard variety \cite{wang2022machine,ramezani2020approaches,bangui2022blockchain} leading to considering new perspectives to provide new self-healing and self-adaptation options.

Thus, our goal in this paper is to advise resilient actors on how to "learn by doing" to reach an acceptable self-adaptation and performance to deal with the continuous improvement of vulnerable systems. Mainly, we focus on examining how to gain from the antifragility concept to instruct resilient actors on improving system-level qualities.
The actor model has acquired popularity in the serverless cloud and edge computing domains~\cite{barcelona-pons_faas_2019,sreekanti_cloudburst_2020}. In this paper, we argue about the importance of supervision trees and custom strategies for reaching the requirements of serverless antifragile systems. We put forward the following contributions:
\begin{itemize}
   \item we describe the importance of actor models for reaching antifragility of software systems -- in particular, the adoption of supervision trees;
    \item we introduce custom strategies that can be applied for customizing the lifecycle of actors to increase the antifragility of software systems;
\end{itemize}

This paper is structured as follows. In Section \ref{sec:background}, we provide the main background regarding the actor model, supervision trees, and serverless systems. In Section \ref{sec:supervision}, we provide the main contribution as the proposal of antifragile supervision in serverless systems and we mention the plan of action for the proof-of-concept implementation and validation of the approach. In Section~\ref{sec:rel-works}, we discuss the related works in the context of actor models and serverless architectures. In Section~\ref{sec:conclusion}, we provide the main conclusions.

\section{\uppercase{Background}}
\label{sec:background}

\subsection{Actor Model}

The actor model is a mathematical model of concurrent computation with roots dating back to 1973. It was introduced by Hewitt et al.~\cite{hewitt1973session} and used as a model for the theoretical understanding of concurrent computing. The model inspired many practical languages and frameworks in the past. It is again starting to receive significant attention due to the demands of high-throughput and low-latency applications in the times of serverless systems~\cite{taibi2020serverless}.

The system using an actor model consists of location-transparent actors, seen in the model as the universal primitives of concurrent computations. Each actor receives input and responds by:

\begin{enumerate}
    \item sending a finite number of messages to the other actors,
    \item creating a finite number of child actors,
    \item modifying its internal state.
\end{enumerate}

Messages are immutable and exchanged between the actors in an asynchronous way only. Each actor is assigned a mailbox address, which serves as a queue for incoming messages, ensuring that each actor processes only one message at a time. These principles are based on \textit{shared-nothing} architecture~\cite{stonebraker1986case}, which, apart from other benefits, simplifies the programming model by introducing a lock-free development environment.

\begin{figure}[!t]
\centering
\includegraphics[width=2in]{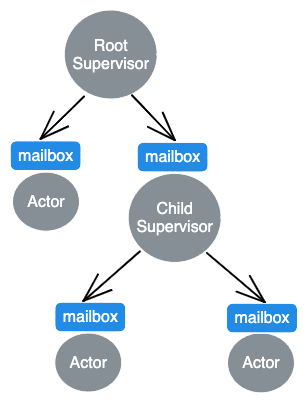}
\caption{Parent-child supervision hierarchy in the actor model}
\label{fig:actor}
\end{figure}

\subsection{Supervision Tree}

The creation of child actors in the actor model forms a hierarchical structure with a single root actor. This resulted in the idea of parent-child supervision~(Fig.~\ref{fig:actor}), which Ericsson popularised as part of the Open Telecom Platform (OTP). OTP includes several ready-to-use components and design principles, which are nowadays integrated into the Erlang/OTP ecosystem. It was that utilisation of Erlang/OTP that helped Ericsson to build their highly available telephony network with reported nine nines availability~\cite{armstrong_concurrency_2007}. Since the actors are standalone distributed instances, the supervision fundamentally differs from the traditional single-call stack runtimes. Such runtimes were designed in the era of single-core machines and came with the illusion of the shared call stack, causing many conceptual problems when used with concurrent models~\cite{documentation_actor_nodate}. Nevertheless, supervision has become a tool to embrace failure~\cite{boner2014reactive} and was integrated into mature actor-model languages and frameworks, such as Erlang, Elixir and Akka.

Supervision enables to push error prone functionality to the leaf actors and lets them crash in case of unexpected failures. In case of such failure, it is up to the supervising parent to implement a strategy to mitigate it. Based on the configured strategy, the parent then performs a supervision directive to either restart, resume, or stop the problematic child actor. In case of lacking knowledge or competence, the supervising parent can escalate the issue further up the tree.

The \textit{let it crash} principle enables the developer to achieve a more readable offensive code style~\cite{noauthor_offensive_nodate} without worrying about influencing the rest of the tree in case of unexpected failures. Such unexpected failures are known as transient Heisenbugs~\cite{gray_why_1985}, and the supervisor can react to them based on the configured strategies and directives, leading to the self-healing of the actor instances and overall resiliency of the whole tree. On the other hand, expected failures can be treated as any other domain events with simple message passing, which can be enhanced even by adopting the Railway Oriented Programming (ROP) style~\cite{eason2018railway}.

\subsection{State of Serverless}

In serverless systems, the code is executed in stateless containers triggered by events and structured as Functions as a Service (FaaS)~\cite{taibi2020serverless}. In FaaS, each function can represent a small part of the application. Differently from services in a Microservices environment, the functions have a limited time span when they are instantiated on-demand~\cite{taibi2020serverless}.

Based on the analysis of the available serverless applications~\cite{eismann_state_2022}, up to 61\% of the applications rely on some form of underlying storage holding the application state.

Amazon Web Services (AWS), with its AWS Lambda services, dominate the serverless Function as a Services (FaaS) cloud provider platforms, taking the majority of the market share ~\cite{eismann_state_2022,noauthor_stack_nodate}. Although most applications rely on the application state, FaaS platforms usually depend on the stateless nature of functions. For example, constructing a stateful application with AWS Lambda requires the coupling of functions with external storage services. For these cases, Amazon offers highly available and highly scalable storage services, such as Amazon DynamoDB~\cite{decandia_dynamo_nodate} for key-value storage or AWS Simple Storage Service (S3) for object storage.

Scaling or self-healing of stateless functions is a relatively easy task due to their idempotency, as discussed by many previous studies~\cite{helland_toobig_2017,castro_rise_2019}. Horizontal scaling can be achieved by adding more services with a load balancer in front. Self-healing, on the other hand, usually means a simple restart without worrying about side effects and about losing any current state. Stateless functions, however, defer the complex state-handling logic, such as scaling writes, to the developers. Scaling writes, especially in distributed systems, is far more challenging than scaling reads since one usually wants to achieve at least a reasonable eventual consistency~\cite{vogels_eventually_2009} without worrying about concurrent access to the same resource. After applying practices such as data sharding or Command Query Responsibility Segregation (CQRS), the next viable option is to rely on the underlying storage support for optimistic locking~\cite{halici_optimistic_1991}, which is only feasible until concurrent access is encountered relatively rarely. In case of a high probability of concurrent access, it is up to the developer to achieve a Single Writer Principle (SWP)~\cite{thompson_mechanical_2011}, which is technically a very complex task to achieve in a distributed environment~\cite{ludwikowski_when_2021}.

\subsection{Serverless Actors}
Actors, on the other side, provide a similar level of granularity as stateless functions, but compared to functions, actors are stateful by default. In addition, the infrastructure for self-healing or scaling is often built-in inside the existing robust actor model-based ecosystems. For example, Akka, Lightbend's actor model-based framework, supports the Distributed SWP as part of the Akka Cluster Sharding module~\cite{ludwikowski_when_2021,enes_single-writer_2017} with the possibility of strong consistency~\cite{lightbend_distributed_2022} according to CAP theorem~\cite{gilbert02cap}. Combined with fast append-only event-sourced persistence provided by the Akka Persistence module, most of the highly complex but common infrastructural issues are provided at the framework's level.

Providing similar functionalities inside the serverless environments, combined with sub-second billing and the potential for infinite scalability of actors, the actor model can be a viable option for writing stateful serverless applications. Compared to the stateless functions, the actor model has the potential to abstract the necessary infrastructure, such as self-healing and scaling, even further, putting the main focus on writing solely what actually matters -- the domain logic.

\subsection{Serverless Platforms}

Microsoft Azure, the second most used FaaS cloud provider platform~\cite{noauthor_stack_nodate}, introduced Reliable Actors~\cite{tomvcassidy_service_nodate} built on top of the stateful Durable Functions~\cite{burckhardt_durable_2021} as part of their Service Fabric Platform as a Service (PaaS). Originating from Microsoft Research on the Orleans~\cite{bernstein_orleans_nodate} project, Reliable Actors bring virtual actors into the serverless environment. 
Furthermore, the Reliable Actors are further utilized in serverless runtime environments, such as Microsoft's Distributed Application Runtime (Dapr).

WasmCloud is another platform utilizing the serverless actor model. The projects aim to develop applications in WebAssembly, without an infrastructural boilerplate. Individual actors are the smallest deployable units in the cloud, which facilitates microservices-like deployment. Due to the low memory footprint runtime of WebAssembly, actors are well-suitable for deployment into a cluster at the edge. Orchestration itself can be potentially utilized with the help of Kubernetes, by adopting Krustlet Kubelet~\cite{rac_at_2021}.

Cloudflare Workers represent another viable option for writing serverless actors, as they support the actor model through their Durable Objects~\cite{varda_workers_2020}. Similarly to WasmCloud and $\mu$Actor, they rely on a low memory footprint runtime, promoting the suitability for real-time applications and edge computing via distributed databases at the edge.

Hetzel et al.~\cite{hetzel_actor_2021} proposed a stateful platform called $\mu$Actor by utilizing an actor model that can run in the whole edge-cloud continuum. They specifically focused on the microcontrollers in the edge computing domain, which are not able to run heavyweight virtual machines or even containers. Instead, they rely on a low-memory footprint runtime running Lua, which does not cause cold start issues and neither prevents running operations in the leaf nodes. 
To sum up, existing platforms are focused on proposing resilient solutions to discover and prevent the root causes of vulnerabilities. However, existing resilient solutions focus mainly on helping serverless applications recover from the negative impacts, having yet to address self-adaptability learning from runtime events.

\section{\uppercase{Antifragile Serverless Supervision}}
\label{sec:supervision}

Even though the actor model is nowadays becoming popular in the serverless cloud and edge computing domains, the existing serverless actor-model-based frameworks do not support the creation of supervision trees. This is either due to the low maturity of the existing frameworks or due to the frameworks being based on the virtual actors, which are immediately re-instantiated on failure by the runtime -- without a possibility of applying any kind of supervision strategy. Automatic restart of virtual actors certainly promotes some level of resiliency but does not fit into the antifragile view since the system just tolerates failures instead of utilizing them to improve further. 

\subsection{Built-in strategies}

We support the idea that the ability to configure strategies and directives should not be restricted to serverless developers. Therefore moving the supervision strategies~(Fig.~\ref{fig:built-in-custom}) from the existing robust actor-based ecosystems (Erlang, Akka) into the serverless environment could result in higher resiliency at the expense of a higher development complexity.

\subsection{Custom strategies}

\begin{figure}[!h]
\centering
\includegraphics[width=2.2in]{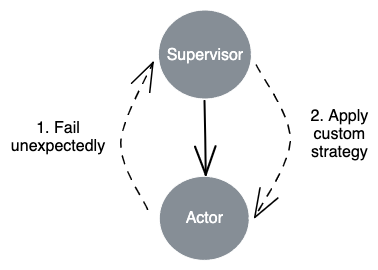}
\caption{Supervising actor lifecycle strategy}
\label{fig:built-in-custom}
\end{figure}

Existing built-in strategies are formed around resuming, restarting, and stopping the given actor with or without its siblings and usually do not offer any further customization~\cite{lightbend_supervision_nodate,ericsson_supervision_nodate}. However, extending supervision by enabling a definition of custom strategies~(Fig.~\ref{fig:built-in-custom}) could be another step towards the high availability and overall resiliency of the serverless actors. In the end, some of the custom but generic enough strategies could be implemented back in the serverless services.

There is also potential for defining strategies based on the expected \textit{domain events}. Handling of expected domain events, including domain errors, could potentially benefit from similar directives applied in case of unexpected failures. Moreover, analyzing the expected domain events can be a step towards preventing failure by applying a set of given preemptive strategies before a potential failure could happen.

In case of independently deployable versioned actors, a type of custom strategy can be based on providing a fallback functionality and falling back to a previous working version of the failing actor. The parent supervisor could spawn other actors, which could either try to heal the failed actor or provide similar functionality but in a less error-prone way.

Strategies can also be applied for \textit{orchestration}, ranging from load-balancing of stateless actors to deployment strategies based on heuristics\cite{tardieu_reliable_2022} for individual actors. Similarly to the orchestration of the FaaS, containers orchestration based on Kubernetes, or scaling out/in policies based on the resource metrics, actors need the orchestration strategies provided both on the infrastructural and on the application level.

\subsection{Antifragile Strategies}

\begin{figure}[!h]
\centering
\includegraphics[width=3in]{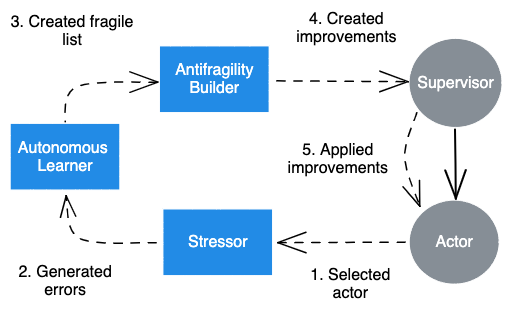}
\caption{Predictive strategies}
\label{fig:predictive}
\end{figure}

Custom supervision strategies have the potential to be implemented based on the predictive model, which could result in a more robust system in response to negative incidents - an antifragile system. A similar idea is proposed in the microservices domain, which is based on applying external stressors to a microservice~\cite{hind-iotbds22,hind-icsme2022}. In our case~(Fig.~\ref{fig:predictive}), instead of stressing a microservice, we stress an individual actor, consequentially resulting an a more robust version of the actor itself or strengthening the supervisor-child actor relationship.

Stress is the process of intentionally introducing scenarios which could result in the failure of a system component - an actor or a whole tree of actors. The term "intentional stress" means artificially injecting failures into the system or deploying a defective system component and exposing it to a stressful environment. In that sense, stress is related to a state of a system component in both runtime and design time periods.

Actors can be picked to be stressed based on several strategies, ranging from analysing the history of past runtime incidents to choosing actors responsible for the system's critical parts. For the sake of simplicity, we can argue that in the case of limitless resources, stress can be applied to each individual actor in the system.

The whole process of applying antifragile strategy consists of the following components:

\begin{enumerate}
    \item \textbf{Stressor} selects an Actor to stress, on top of which it will try to generate errors,

    \begin{itemize}
        \item The selected actor can be stressed independently or together with its hierarchy.
        \item Stress can be performed in a production environment or a virtual sandbox environment.
        \item Stressor can artificially create stressful situations, i.e. based on the expected non-functional requirements or be represented by production load and cope with stressful situations on demand.
    \end{itemize}

    \item \textbf{Autonomous Learner}, as a machine learning component, is responsible for analysing the generated errors outputting a list of system fragilities,

    \item \textbf{Antifragility Builder}, as another component which is responsible for analysing the fragilities and building a list of antifragile improvements,

    \begin{itemize}
        \item Improvements can be external to the actor and related to the implementation of the supervision strategy or internal, meaning modifying the implementation of the stressed actor.
        \item Application of external and internal improvements can be automatically and gradually distributed by re-deploying new versions of respective actors.
    \end{itemize}

    \item \textbf{Supervisor} is the actor responsible for managing the lifecycle of its child actors throughout the implemented supervision strategies. In case of internal improvements, the supervisor can pass the list of improvements to the child, which could be updated, automatically re-deployed and gradually activated by the supervisor as an improved version of the same actor,

    \item \textbf{Actor} is the selected stressed child actor, which is managed by its supervisor.

\end{enumerate}

\subsection{Action Plan}
\label{sec:plan}
The biggest argument against supervision is the additional complexity of defining different strategies and directives due to managing failed actors. However, the increased software development complexity is an anticipated and accepted metric in all the critical domains. There is already a need for rigorous testing or software verification due to high resiliency demands. This is why the next step in this research is a proof-of-concept solution validating the proposed supervision strategies, which is now ongoing. We plan to validate our ideas by extending the built-in supervision strategies in the existing actor model-based frameworks (i.e., Akka) outside the serverless environment. However, we know that implementing custom strategies with the existing actor-model frameworks is not trivial. Ultimately, we still need to apply these ideas in the serverless environment. Therefore, we intend to focus directly on the available serverless platforms. Since, to our knowledge, there is no out-of-the-box support for the supervision strategies in the current serverless platforms, we plan to explore the following options to validate our ideas:

\begin{enumerate}
    \item extend the existing actor model-based open-source serverless platform with the supervision strategies (i.e. WasmCloud),
    \item create a simple proof-of-concept solution for the actor model and the strategies on top of the existing open-source serverless runtime (i.e. FAASM),
    \item simulate the actor-model strategies using the existing actor-model-based serverless platform (i.e. Cloudflare Durable Objects).
    \item run experiments about the implemented proof-of-concept solution to apply software systems' stressors  by injecting failures at runtime (e.g., by adopting chaos engineering toolkits).
\end{enumerate}

\section{\uppercase{Related Works}}
\label{sec:rel-works}

The self-healing resilience strategy realizes the idea that an actor can always bounce back to the original condition. In contrast, our antifragility strategy requires an actor capable of reaching a previously unexpected condition. Enabling an actor to learn from shocks, random events, or stresses is highly desirable as it allows the continuous improvement of smart environments. For instance, actors with learning abilities have been suggested in \cite{cao2021intelligent} to help the constant observation and optimization of healthcare systems. In this work, the role of actors mainly focuses around how to achieve their tasks to provide an acceptable quality of service. However, considering how actors can gain from faults is not discussed.

Similarly, in other existing studies, e.g., \cite{daniel2018serverless,barcelona-pons_faas_2019}, the idea of exploring the circumstance of risk occurrence is limited to making an actor robust and reliable but not ready to exploit gains to develop a better understanding of the environment and manage a similar risk in the future.

The research by Barcelona-Pons et al.~\cite{daniel2018serverless} modelled an actor model in a serverless environment on top of the existing AWS services. 
The authors proposed a proof of concept solution and compared it to the traditional FaaS stateless model. Apart from some technical challenges, the results brought nearly 6x better performance in favor of stateful actors, primarily because of the caused latency of saving state per request to the external storage in the case of the stateless functions.

Other research by Barcelona-Pons et al.~\cite{barcelona-pons_faas_2019}, and early projects also bet on statefulness in serverless environments and a higher abstraction of the mentioned infrastructural concerns. Lighbend's Kalix stateful serverless PaaS combines years of Lightbend's experience with the actor model on the Akka framework, promising the developers to focus purely on the domain logic. 

Moreover, another recent research~\cite{barcelona-pons_faas_2019,sreekanti_cloudburst_2020} supports the idea of statefulness, with or without the actors, inside serverless environments.

\section{\uppercase{Conclusion}}
\label{sec:conclusion}
In this paper, we proposed adopting the actor model for building antifragile serverless systems. To move towards antifragility, we suggested supervision trees and custom strategies that can be applied for customizing the lifecycle of actors to increase the antifragility of serverless systems. We proposed a predictive strategy based on the concept of stressors, in which actors or a hierarchy of actors can be selected for some stressing activity (i.e., injecting failures). Other components can analyze the behavior of actors and generate a list of system's improvements. In contrast, the supervisor component is responsible for managing the lifecycle of the child actors. Overall, the solution adopting a robust actor model-based ecosystem can improve the system's resiliency in exchange for higher complexity. For this reason, we detailed the next steps for implementing a proof-of-concept solution to evaluate the benefits and limitations of the approach, as we strongly support the adoption of actors models to build antifragile systems.

\section*{Acknowledgement}
The work was supported from ERDF/ESF “CyberSecurity, CyberCrime and Critical Information Infrastructures Center of Excellence” (No. CZ.02.1.01/0.0/0.0/16\_019/0000822).

\bibliographystyle{apalike}
{\small

\begin{thebibliography}{}

\bibitem[Armstrong, 2007]{armstrong_concurrency_2007}
Armstrong, J. (2007).
\newblock Concurrency {Oriented} {Programming} in {Erlang}.
\newblock
  \url{https://www.cs.uml.edu/ecg/uploads/RobotControl2007/Erlang-concurrency-armstrong.pdf}.
\newblock Last accessed on 2022-12-04.

\bibitem[Bangui and B{\"u}hnov{\'a}, 2022]{bangui2022blockchain}
Bangui, H. and B{\"u}hnov{\'a}, B. (2022).
\newblock Blockchain patterns in critical infrastructures: Limitations and
  recommendations.
\newblock {\em Proceedings of the 17th International Conference on Software
  Technologies - Volume 1: ICSOFT,}, pages 457--468.

\bibitem[Bangui. et~al., 2022]{hind-iotbds22}
Bangui., H., Buhnova., B., and Rossi., B. (2022).
\newblock Shifting towards antifragile critical infrastructure systems.
\newblock In {\em Proceedings of the 7th International Conference on Internet
  of Things, Big Data and Security - IoTBDS,}, pages 78--87. INSTICC,
  SciTePress.

\bibitem[Bangui et~al., 2022]{hind-icsme2022}
Bangui, H., Rossi, B., and Buhnova, B. (2022).
\newblock A conceptual antifragile microservice framework for reshaping
  critical infrastructures.
\newblock In {\em International Conference on Software Maintenance and
  Evolution, {ICSME} 2022, Limassol, Cyprus, October 2-7, 2022}. {IEEE}.

\bibitem[Barcelona-Pons et~al., 2018]{daniel2018serverless}
Barcelona-Pons, D., Ruiz, {\'A}., Arroyo-Pinto, D., and Garc{\'i}a-L{\'o}pez,
  P. (2018).
\newblock {S}tudying the feasibility of serverless actors.
\newblock In {\em ESSCA'18}, volume 2330 of {\em CEUR Workshop Proceedings},
  pages 25--29. CEUR-WS.org.

\bibitem[Barcelona-Pons et~al., 2019]{barcelona-pons_faas_2019}
Barcelona-Pons, D., Sánchez-Artigas, M., París, G., Sutra, P., and
  García-López, P. (2019).
\newblock On the {FaaS} {Track}: {Building} stateful distributed applications
  with serverless architectures.
\newblock In {\em Proceedings of the 20th {International} {Middleware}
  {Conference}}, pages 41--54. ACM.

\bibitem[Bernstein et~al., 2014]{bernstein_orleans_nodate}
Bernstein, P., Bykov, S., Geller, A., Kliot, G., and Thelin, J. (2014).
\newblock Orleans: {Distributed} {Virtual} {Actors} for {Programmability} and
  {Scalability}.
\newblock Technical Report MSR-TR-2014-41, Microsoft.

\bibitem[Bon{\'e}r et~al., 2014]{boner2014reactive}
Bon{\'e}r, J., Farley, D., Kuhn, R., and Thompson, M. (2014).
\newblock The reactive manifesto.
\newblock {\em Dosegljivo: http://www. reactivemanifesto. org/.[Dostopano: 21.
  08. 2017]}.

\bibitem[Burckhardt et~al., 2021]{burckhardt_durable_2021}
Burckhardt, S., Gillum, C., Justo, D., Kallas, K., McMahon, C., and Meiklejohn,
  C.~S. (2021).
\newblock Durable functions: semantics for stateful serverless.
\newblock {\em Proceedings of the ACM on Programming Languages},
  5(OOPSLA):1--27.

\bibitem[Cao et~al., 2021]{cao2021intelligent}
Cao, Z., Liu, L., and Wang, J. (2021).
\newblock Intelligent health information services requirements revisited from
  an actor's perspective.
\newblock In {\em 2021 IEEE International Conference on Digital Health (ICDH)},
  pages 244--253. IEEE.

\bibitem[Cassidy, 2022]{tomvcassidy_service_nodate}
Cassidy, T. (2022).
\newblock Service {Fabric} {Reliable} {Actors} {Overview} - {Azure} {Service}
  {Fabric}.
\newblock
  \url{https://learn.microsoft.com/en-us/azure/service-fabric/service-fabric-reliable-actors-introduction}.
\newblock Last accessed on 2022-12-04.

\bibitem[Castro et~al., 2019]{castro_rise_2019}
Castro, P., Ishakian, V., Muthusamy, V., and Slominski, A. (2019).
\newblock The rise of serverless computing.
\newblock {\em Commun. ACM}, 62(12).

\bibitem[Cunningham, 2011]{noauthor_offensive_nodate}
Cunningham (2011).
\newblock Offensive {Programming}.
\newblock \url{http://wiki.c2.com/?OffensiveProgramming}.
\newblock Last accessed on 2022-12-03.

\bibitem[Cutter et~al., 2013]{national2012disaster}
Cutter, S.~L., Ahearn, J.~A., Amadei, B., Crawford, P., Eide, E.~A., Galloway,
  G.~E., Goodchild, M.~F., Kunreuther, H.~C., Li-Vollmer, M., Schoch-Spana, M.,
  et~al. (2013).
\newblock Disaster resilience: A national imperative.
\newblock {\em Environment: Science and Policy for Sustainable Development},
  55(2):25--29.

\bibitem[De~Bleser, 2020]{de2020automated}
De~Bleser, J. (2020).
\newblock {\em An Automated Delta-Debugging Approach to Resilience Testing of
  Actor Systems through Fault Injection}.
\newblock PhD thesis, Vrije Universiteit Brussel.

\bibitem[DeCandia et~al., 2007]{decandia_dynamo_nodate}
DeCandia, G., Hastorun, D., Jampani, M., Kakulapati, G., Lakshman, A., Pilchin,
  A., Sivasubramanian, S., Vosshall, P., and Vogels, W. (2007).
\newblock {Dynamo: Amazon's highly available key-value store}.
\newblock {\em SIGOPS Oper. Syst. Rev.}, 41(6):205--220.

\bibitem[Eason, 2018]{eason2018railway}
Eason, K. (2018).
\newblock Railway oriented programming.
\newblock In {\em Stylish F\#}, pages 283--307. Springer.

\bibitem[Eismann et~al., 2022]{eismann_state_2022}
Eismann, S., Scheuner, J., Eyk, E.~v., Schwinger, M., Grohmann, J., Herbst, N.,
  Abad, C.~L., and Iosup, A. (2022).
\newblock The {State} of {Serverless} {Applications}: {Collection},
  {Characterization}, and {Community} {Consensus}.
\newblock {\em IEEE Transactions on Software Engineering}, 48(10):4152--4166.

\bibitem[Enes et~al., 2017]{enes_single-writer_2017}
Enes, V., Almeida, P.~S., and Baquero, C. (2017).
\newblock The {Single}-{Writer} {Principle} in {CRDT} {Composition}.
\newblock In {\em Proceedings of the {Programming} {Models} and {Languages} for
  {Distributed} {Computing}}, {PMLDC} '17, pages 1--3, New York, NY, USA.
  Association for Computing Machinery.

\bibitem[Ericsson, 1999]{ericsson_supervision_nodate}
Ericsson (1999).
\newblock Supervision {Principles}.
\newblock
  \url{https://erlang.org/documentation/doc-4.9.1/doc/design_principles/sup_princ.html}.
\newblock Last accessed on 2022-12-04.

\bibitem[Gilbert and Lynch, 2002]{gilbert02cap}
Gilbert, S. and Lynch, N. (2002).
\newblock Brewer's conjecture and the feasibility of consistent, available,
  partition-tolerant web services.
\newblock {\em SIGACT News}, 33(2):51--59.

\bibitem[Gray, 1986]{gray_why_1985}
Gray, J. (1986).
\newblock Why do computers stop and what can be done about it?
\newblock In {\em Symposium on reliability in distributed software and database
  systems}, pages 3--12. Los Angeles, CA, USA.

\bibitem[Halici and Dogac, 1991]{halici_optimistic_1991}
Halici, U. and Dogac, A. (1991).
\newblock An optimistic locking technique for concurrency control in
  distributed databases.
\newblock {\em IEEE Transactions on Software Engineering}, 17(7):712--724.

\bibitem[Helland et~al., 2017]{helland_toobig_2017}
Helland, P., Weaver, S., and Harris, E. (2017).
\newblock Too big not to fail: Embrace failure so it doesn’t embrace you.
\newblock {\em Queue}, 15(1):57–70.

\bibitem[Hetzel et~al., 2021]{hetzel_actor_2021}
Hetzel, R., Kärkkäinen, T., and Ott, J. (2021).
\newblock $\mu$actor: {Stateful} {Serverless} at the {Edge}.
\newblock In {\em Proceedings of the 1st {Workshop} on {Serverless} mobile
  networking for {6G} {Communications}}, pages 1--6, Virtual WI USA. ACM.

\bibitem[Hewitt et~al., 1973]{hewitt1973session}
Hewitt, C., Bishop, P., and Steiger, R. (1973).
\newblock Session 8 formalisms for artificial intelligence a universal modular
  actor formalism for artificial intelligence.
\newblock In {\em Advance Papers of the Conference}, volume~3. Stanford
  Research Institute Menlo Park, CA.

\bibitem[Lightbend, 2022a]{documentation_actor_nodate}
Lightbend (2022a).
\newblock Actor model :: {Akka} {Guide}.
\newblock
  \url{https://developer.lightbend.com/docs/akka-guide/concepts/akka-actor.html}.
\newblock Last accessed on 2022-12-03.

\bibitem[Lightbend, 2022b]{lightbend_distributed_2022}
Lightbend (2022b).
\newblock Distributed {Data} • {Akka} {Docs}.
\newblock
  \url{https://doc.akka.io/docs/akka/current/typed/distributed-data.html#write-consistency}.
\newblock Last accessed on 2022-11-22.

\bibitem[Lightbend, 2022c]{lightbend_supervision_nodate}
Lightbend (2022c).
\newblock Supervision and {Monitoring} • {Akka} {Docs}.
\newblock \url{https://doc.akka.io/docs/akka/2.5/general/supervision.html}.
\newblock Last accessed on 2022-12-04.

\bibitem[Ludwikowski, 2021]{ludwikowski_when_2021}
Ludwikowski, A. (2021).
\newblock When do you need {Akka} {Cluster}?
\newblock
  \url{https://blog.softwaremill.com/when-do-you-need-akka-cluster-5885d43e901b}.
\newblock Last accessed on 2022-11-22.

\bibitem[Overflow, 2022]{noauthor_stack_nodate}
Overflow, S. (2022).
\newblock Stack {Overflow} {Developer} {Survey} 2022.
\newblock
  \url{https://survey.stackoverflow.co/2022/?utm\_source=social-share&utm\_medium=social&utm\_campaign=dev-survey-2022}.
\newblock Last accessed on 2022-11-22.

\bibitem[Rac and Brorsson, 2021]{rac_at_2021}
Rac, S. and Brorsson, M. (2021).
\newblock At the {Edge} of a {Seamless} {Cloud} {Experience}.
\newblock arXiv:2111.06157 [cs].

\bibitem[Ramezani and Camarinha-Matos, 2020]{ramezani2020approaches}
Ramezani, J. and Camarinha-Matos, L.~M. (2020).
\newblock Approaches for resilience and antifragility in collaborative business
  ecosystems.
\newblock {\em Technological Forecasting and Social Change}, 151:119846.

\bibitem[Sreekanti et~al., 2020]{sreekanti_cloudburst_2020}
Sreekanti, V., Wu, C., Lin, X.~C., Schleier-Smith, J., Faleiro, J.~M.,
  Gonzalez, J.~E., Hellerstein, J.~M., and Tumanov, A. (2020).
\newblock Cloudburst: {Stateful} {Functions}-as-a-{Service}.
\newblock {\em Proceedings of the VLDB Endowment}, 13(12):2438--2452.
\newblock arXiv:2001.04592 [cs].

\bibitem[Stonebraker, 1986]{stonebraker1986case}
Stonebraker, M. (1986).
\newblock The case for shared nothing.
\newblock {\em IEEE Database Eng. Bull.}, 9(1):4--9.

\bibitem[Taibi et~al., 2020]{taibi2020serverless}
Taibi, D., Spillner, J., and Wawruch, K. (2020).
\newblock Serverless computing-where are we now, and where are we heading?
\newblock {\em IEEE software}, 38(1).

\bibitem[Taleb, 2012]{taleb2012antifragile}
Taleb, N.~N. (2012).
\newblock {\em Antifragile: Things that gain from disorder}, volume~3.
\newblock Random House Incorporated.

\bibitem[Tardieu et~al., 2022]{tardieu_reliable_2022}
Tardieu, O., Grove, D., Bercea, G.-T., Castro, P., Cwiklik, J., and Epstein, E.
  (2022).
\newblock Reliable {Actors} with {Retry} {Orchestration}.
\newblock arXiv:2111.11562 [cs].

\bibitem[Thompson, 2011]{thompson_mechanical_2011}
Thompson, M. (2011).
\newblock Mechanical {Sympathy}: {Single} {Writer} {Principle}.
\newblock
  \url{https://mechanical-sympathy.blogspot.com/2011/09/single-writer-principle.html}.
\newblock Last accessed on 2022-11-22.

\bibitem[Varda, 2020]{varda_workers_2020}
Varda, K. (2020).
\newblock Workers {Durable} {Objects} {Beta}: {A} {New} {Approach} to
  {Stateful} {Serverless}.
\newblock
  \url{http://blog.cloudflare.com/introducing-workers-durable-objects/}.
\newblock Last accessed on 2022-12-04.

\bibitem[Vogels, 2009]{vogels_eventually_2009}
Vogels, W. (2009).
\newblock Eventually consistent.
\newblock {\em Communications of the ACM}, 52(1):40--44.

\bibitem[Wang et~al., 2022]{wang2022machine}
Wang, X., Mazumder, R.~K., Salarieh, B., Salman, A.~M., Shafieezadeh, A., and
  Li, Y. (2022).
\newblock Machine learning for risk and resilience assessment in structural
  engineering: Progress and future trends.
\newblock {\em Journal of Structural Engineering}, 148(8):03122003.

\end{thebibliography}

}
\balance

\end{document}